\documentclass[twocolumn,twoside,slac_two]{revtex4}
\usepackage{graphicx}
\usepackage{fancyhdr}

\usepackage[english]{babel}

\usepackage{hyperref}
\hypersetup{
    colorlinks=true,
    urlcolor=magenta,
}
\begin{document}

\title{Calculations of CR energy spectra within the NoRD model}

%

\author{V.V. Uchaikin, R.T. Sibatov}
\affiliation{Ulyanovsk State University, 42 Leo Tolstoy street, Ulyanovsk, Russia}

\begin{abstract}

Energy spectra of galactic cosmic rays calculated within the framework of the NoRD (nonlocal relativistic diffusion) model are presented.
The model accounts for the turbulent character of the interstellar medium and the relativistic speed limit requirement. Calculations account for spiral distribution of sources, boundedness of halo, spallation of nuclei, energy dependence of the diffusion coefficient, tempered power law injection spectrum. We show that the knee in the background spectra from the ensemble of supernovae can arise due to relativistic speed limit requirement. We compare our calculations with the results obtained in frames of the local models (LoD and LoRD) and nonlocal nonrelativistic model (NoD).

\end{abstract}

\maketitle

\thispagestyle{fancy}


\section{Introduction}

The first explanation (or rather prediction) of the power law of primary cosmic ray spectrum made by E. Fermi was purely phenomenological: it was based on the exponential growth of energy with time and exponential distribution of ages of observed particles. Both of them ignored the real space-time distribution of sources, highly inhomogeneous structure of the interstellar medium and boundedness of the Galaxy. A few years later, the CR propagation was interpreted as some kind of Brownian  motion and began to be calculated in terms of the ordinary (normal) diffusion theory \cite{Ginzburg}. It is easy to understand the origin of this model: Fermi wrote about random walk of CR particles among magnetic clouds chaotically distributed in interstellar space, and Brownian motion was the most expressive image of such a process. However, its mathematical representation in the form of the \textit{Local Diffusion} (LoD) equation (or its other local modifications) did not involve the particle velocity in an explicit form and for this reason failed in description of front propagation and some other pecularities of the CR transport. This is just the same task which inspired authors of the work \cite{Lagutin01} to seek a new propagator in the frame of \textit{Nonlocal Diffusion} (NoD) model.

We talk here on locality because the generic sign of the diffusion approximation is the local Laplace operator $\Delta\equiv\nabla^2=\frac{\partial^2}{\partial x^2}+
\frac{\partial^2}{\partial y^2}+
\frac{\partial^2}{\partial z^2}$. Its application assumes uniformity and mutually independence of points of scattering like those we observe in ideal gases. No doubt, distribution of irregularities in the interstellar magnetic field do not possess this property, they are linked with each other by means of more or less long magnetic force lines. These lines are frozen into the material component of interstellar plasma and are involved into the common turbulent motion. It is rather strange, why the results in turbulent diffusion theory obtained in classical works followed the Kolmogorov (1941) phenomenological basis of the turbulence \cite{Weiz,Heisenberg,Tchen,Monin55}
were completely ignored in the early development of the CR-propagation theory. One of the generic features of the turbulent diffusion is the presence of nonlocal operators in its equation.

Involvement of non-local operator $\Delta^{\alpha/2}$ instead of its local counterpart $\Delta$ has shown that in case of a single local pulse source the knee arises without any additional assumptions except the turbulent (called also \textit{fractal}) character of the interstellar medium. During 2001-2010 years, a series of calculations were performed by Lagutin et al. with the use of NoD-model, but all results were obtained without account for the relativistic limitation of speed and the boundary conditions. The speed limitation was inserted into the NoD-model in \cite{Uchaikin10, Uchaikin12} and this version was later named the \textit{Nonlocal Relativistic Diffusion} (NoRD) model \cite{Sibatov15, Erlykin15}.
The physical interpretation of the NoRD model and first results obtained in its frame were described in reviews \cite{Uchaikin13, Uchaikin15}. Here, we report some new results of numerical investigations in the frame of this model adding account of multiplicity of sources and boundedness of halo.

\section{Sources}

Modeling of CR sources was performed under following assumptions.
\begin{enumerate}
  \item  All sources are isotropic, point and instantaneous with equal powers and energy spectra $S_k(E)\sim E^{-\gamma}\exp(-E/E_{\mathrm{max},k})$. Here $\gamma$ is an exponent of injection spectrum, and $E_{\mathrm{max},k}$ characterizes maximum energies of $k$-type nuclei accelerated at supernova remnants.
  \item The sources are separated into two groups: relatively young supernovae (less than ${1.5\cdot 10^6}$ years) whose space-time coordinates were taken individually (15 objects: Vela, Monogem Ring, Geminga and others), and 30000 sources with spatial distribution generated according to approach proposed in \cite{Faucher-Giguere} and used in \cite{Blasi1}. The radial distribution of discrete sources was simulated according to the density of pulsars (see \cite{CaseBhattacharya}),
            $$
            f(r)=\frac{\beta^4 e^{-\beta}}{12\pi R^2_\odot}\left(\frac{r}{R_\odot}\right)^2\exp\left(-\beta\frac{r-R_\odot}{R_\odot}\right)
        $$
with $R_\odot=8.5$~kpc and $\beta=3.53$, the
$|z|$-distribution was modeled by exponential function with $z_g= 200$~pc.
To simulate the spiral structure, the algorithm proposed in~\cite{Faucher-Giguere} was used.
\end{enumerate}

\begin{figure}[tbh]
\centering
\includegraphics[width=0.45\textwidth]{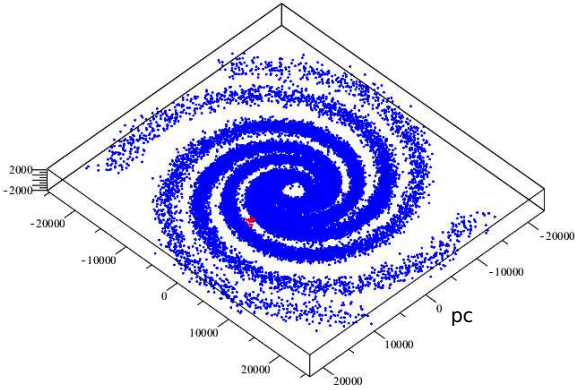}
\caption{Spatial arrangement of sources (red points are known nearest sources, blue points denote distant sources generated by Monte Carlo algorithm).}\label{fig_sources}
\end{figure}

The example realization is presented in Fig.~\ref{fig_sources}.



\section{Propagators}

The area of CR diffusion is the cylinder of height $2H=8$~kpc with radius $R_d=20$~kpc
surrounded by a non-reflective environment. The galactic disk being in the middle of the cylinder parallel to bases has the thickness $2h=600 pc$.
As in Ref.~\cite{Blasi1}, we assume that the diffusion coefficient is independent of spatial and time variables, and
determined by particle rigidity $R$,
\begin{equation}\label{eq_diffusion_coefficient}
D(E)= D_0 (R /3\ GV)^\delta\ [\mathrm{pc}^2 \mathrm{yr}^{-1}], \quad R>3GV.
\end{equation}
Results reported here are related to values $\delta=0.6$ and $D_{0}=0.0729$~$\mathrm{pc}^2 \mathrm{yr}^{-1}$. 

The simplest propagator of the LoD-model in the infinite homogeneous space has the Gaussian form
$$
G_\mathrm{LoD}(\mathbf{r},t, E; \mathbf{r}_s, t_s, E_0)=
$$
\begin{equation}\label{eq_LoD}
=\frac{\delta(E-E_0)}{(4\pi D(E) \tau)^{3/2}}\exp\left(-\frac{(\mathbf{r}-\mathbf{r}_s)^2}{4 D(E) \tau}\right);\quad \tau=t-t_s,
\end{equation}

To account for the halo boundedness, solutions (\ref{eq_LoD}) should be modified: considering zero boundary conditions and neglecting by lateral surface of the cylinder, the image method can be used in local approach,
$$
G_\mathrm{LoD,H}(\mathbf{r},t, E; \mathbf{r}_s, t_s, E_0)=
$$
\begin{equation}\label{eq_LoD_halo}
=\frac{\delta(E-E_0)}{(4\pi D(E) \tau)^{3/2}}\exp\left(-\frac{(x-x_s)^2+(y-y_s)^2}{4 D(E) \tau}\right)\times
$$
$$
\times\sum_{j=-\infty}^{\infty}(-1)^j\exp\left(-\frac{(z-z_j)^2}{4D(E)\tau}\right).
\end{equation}
Here $z_j=(-1)^j z_s+2jH$ are $z$-coordinates of the image sources, $\{x_s,y_s,z_s\}$ are coordinates of an original source.

The NoD-propagator in an infinite medium is expressed through fractional stable L\'evy-Feldheim density \cite{Lagutin01}. In case of $\beta=1$ it can be presented in the form \cite{UchSib12}
{\small
\begin{equation}\label{eq_NoD}
G_\mathrm{NoD}(\mathbf{r},t, E; \mathbf{r}_0, t_0, E_0)=
\end{equation}
$$
=\int\limits_0^\infty G_\mathrm{LoD}(\mathbf{r},\theta, E; \mathbf{r}_0, 0, E_0) \frac{g_+(\theta (t-t_0)^{-2/\alpha};\alpha/2)}{(t-t_0)^{2/\alpha}}d\theta.
$$
}
Here $g_+(x,\nu)$ is a one-sided L\'evy stable density (subordinator, $0<\nu\leq1$) and  $G_\mathrm{LoD}$ in (\ref{eq_NoD}) be taken with  the diffusion coefficient linked with $D_\alpha$ of the NoD-model by relation
\begin{equation}\label{eq_diff_coeff}
D(E)=[D_\alpha(E)]^{2/\alpha}.
\end{equation}
If $D(E)\propto E^\delta$, then $D_\alpha(E)\propto E^{\delta_\alpha}$, where $\delta_\alpha=\delta\alpha/2$. For example, for $\delta=0.6$ and $\alpha=1.5$, we have $D_\alpha(E)\propto E^{0.45}$.

Formula (\ref{eq_NoD}) is suitable for Monte Carlo representation of the NoD-propagator
\begin{equation}\label{eq_NoD_MC}
   G_\mathrm{NoD}=\left\langle G_\mathrm{LoD}(\mathbf{r},\Theta, E; \mathbf{r}_0, 0, E_0) \right\rangle_{\Theta_{\alpha/2}(t-t_0)},
\end{equation}
where $\Theta_{\alpha/2}$ is a stable random variable generated according to (see e.g. \cite{Stable}) $\Theta_{\alpha/2}(t-t_0)=$
$$
=(t-t_0)^{2/\alpha}\frac{\sin(\alpha\pi U_1/2)[
\sin((1-\alpha/2)\pi U_1)]^{2/\alpha-1}}{[\sin (\pi U_1)]^{2/\alpha}[\ln
U_2]^{2/\alpha-1}}.
$$
Formulae (\ref{eq_diff_coeff}) and (\ref{eq_NoD_MC}) indicate the correspondence principle between NoD- and LoD-models (when $\alpha\rightarrow2$) and allow us to use the standard diffusion coefficient to estimate CR characteristics in frames of the nonlocal models.

In calculations of energy spectrum in frames of the NoRD-model (for $\alpha\in(1,2)$) we have used approximate expression for propagators
$$
G_\mathrm{NoRD}(\mathbf{r},t, E; \mathbf{r}_s, t_s, E_0)
$$
$$
=G_\mathrm{NoD}(\mathbf{r},t, E; \mathbf{r}_s, t_s, E_0)\ 1\left(\frac{}{}c\tau-|\mathbf{r}-\mathbf{r}_s|\right)
$$
\begin{equation}\label{eq_NoRD}
+\delta(E-E_0)\delta(|\mathbf{r}-\mathbf{r}_s|-c\tau)\mathrm{Prob}(|\mathbf{r}-\mathbf{r}_s|>c\tau),
\end{equation}
where $1(x)$ is the Heaviside step function.

It should be noted
that zero boundary conditions for nonlocal models are fulfilled approximately, if the mean scattering length of CRs much smaller than thickness of the halo.
To account for restriction of the diffusion domain (halo), the generalized Marshak boundary conditions can be used.
The solutions have been modified by means of the image method.
In this approximation, we use $G_\mathrm{LoD,H}$ instead of $G_\mathrm{LoD}$ in expressions~(\ref{eq_NoD}) and (\ref{eq_NoD_MC}) using procedure (\ref{eq_NoRD}).

\begin{figure}[tbh]
  \includegraphics[width=0.47\textwidth]{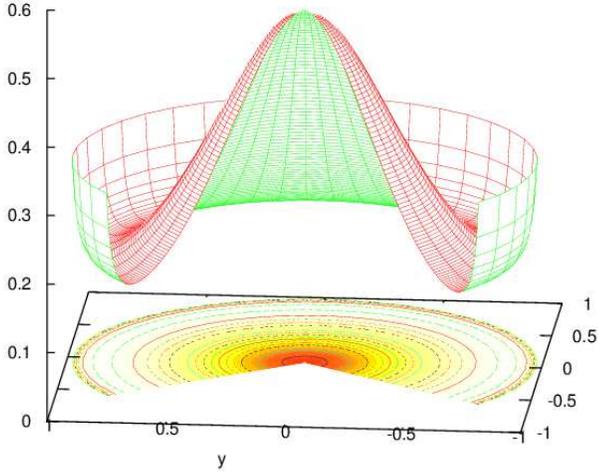}\\
  \caption{Typical shape of the NoRD-propagator in two dimensions.}\label{prop}
\end{figure}

We estimated energy spectra accounting for spatial and temporal distribution of supernova remnants and diffusion in halo in frames of four models (LoD, LoRD, NoD, NoRD). Note, that such calculations for the local model were performed in many papers (see e.g.  \cite{Sveshnikova03, Blasi1, Gaggero}) and for nonlocal model in Refs.~\cite{Lagutin01, Erlykin03, Lagutin15}. All of them, however, utilize propagators ignoring the relativistic speed limit requirement.
Relativistic nature of CRs  sufficiently modify the propagators and spectra as well. NoRD-propagators are restricted to domain $|\mathbf{r}-\mathbf{r}_s|\leq vt$, have ballistic splash at this boundary whereas in the middle part they are spreading superdiffusively (remind, that we restrict ourselves by values of superdiffusion parameter $\alpha\in(1,2]$).
Spallation effects are taken into account according to Blasi and Amato \cite{Blasi1} using formulae provided in \cite{Horandel}.



 \begin{figure}[tbh]
\centering
\includegraphics[width=0.35\textwidth]{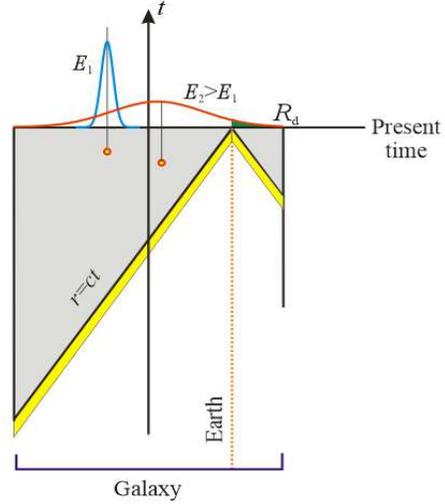}
\caption{To explanation of the steepening in equilibrium spectrum in frames of the LoRD and NoRD model.}\label{fig_steepening}
\end{figure}

\begin{figure}[tbh]
  \includegraphics[width=0.47\textwidth]{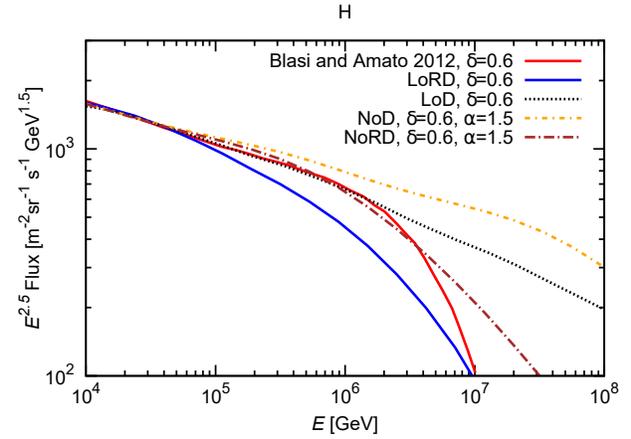}\\
  \caption{Energy spectra for protons calculated in frames of four models of CR diffusion ($\delta=0.6$, $\gamma=2.07$).}\label{proton}
\end{figure}

\begin{figure}[tbh]
  \includegraphics[width=0.47\textwidth]{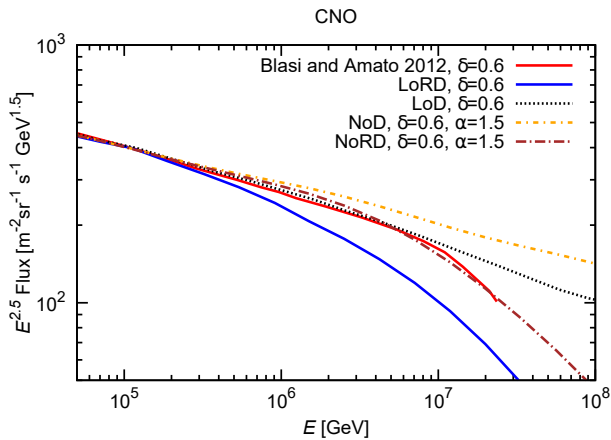}\\
  \caption{Energy spectra for CNO group obtained in frames of four models of CR diffusion ($\delta=0.6$, $\gamma=2.07$).}\label{CNO}
\end{figure}

\begin{figure*}[tbhp]
  \includegraphics[width=0.7\textwidth]{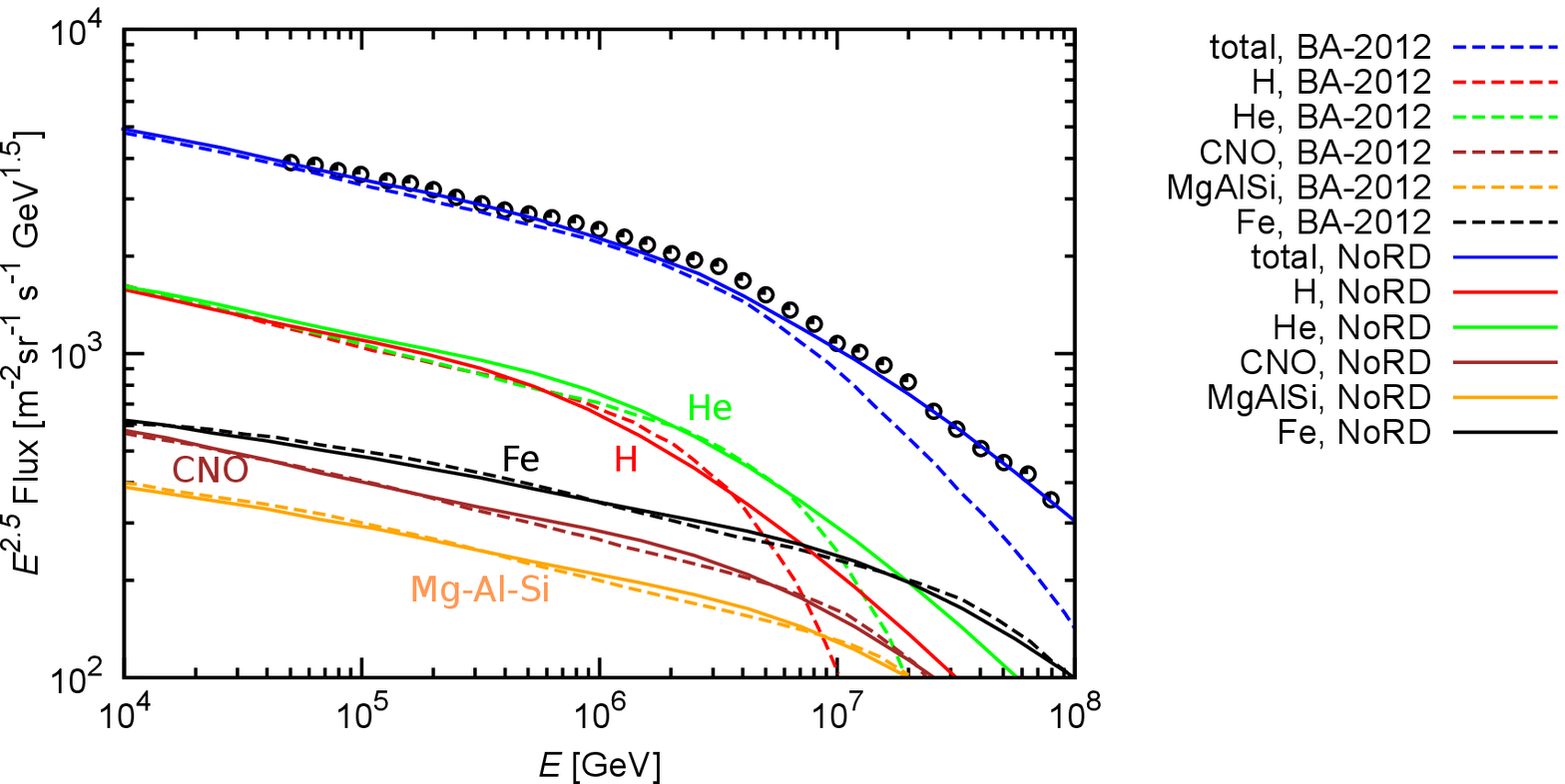}\\
  \caption{Energy spectra in the NoRD model ($\alpha=1.5$, $\delta=0.6$, $\gamma=2.07$) compared with results of Blasi \& Amato (2012).}\label{Total}
\end{figure*}

As is known, the standard assumptions about the origin of galactic CRs: power law spectra of particles injected by SNR, and  diffusive propagation throughout the Galaxy with a coefficient
$D(E)\propto E^\delta$ leads to the spectrum $n(E)\propto N(E)/D(E)\propto E^{-\gamma+\delta}$ (from the SNR ensemble) reflecting the balance between injection and escape of CRs from the confinement volume.
Accounting for relativistic restriction (in terms of relativistic propagators) steepens the spectrum at high energies, and the position of the spectrum break, its sharpness and slopes of asymptotes depend on $D_\alpha(E)$ and~$\alpha$ \cite{Sibatov15}.
The cause of this effect is universal and must be kept in mind in all calculations of propagation of high energy CRs. Explanation of the spectrum steepening at high energies in the NoRD model is schematically represented in Fig.~\ref{fig_steepening}. The grey field corresponds to the space-time area containing CR sources which do not take part in the formation of the spectrum due to the relativistic restriction. Their accounting in the ordinary diffusion model leads to a pure power law spectrum $\propto E^{-\gamma-\delta}$ under the assumption about the pure power law spectrum of injected particles. Taking the relativistic restriction into account in the NoRD model excludes automatically these sources and this exclusion affects predominantly on high energy part of CRs due to $D(E)\propto E^\delta$.

It is important to emphasize that, the NoRD model (compared to the NoD-version used by Lagutin et al.) operates with acceptable values of injection exponent $\gamma$ ($\approx2$)  (in contrast with $\gamma>2.7$, e.g. $\gamma=2.85$, in recent calculations by Lagutin et al.~\cite{Lagutin15}).

Fig.~\ref{proton}, \ref{CNO} and \ref{Total} demonstrate computational results for spectrum of protons, CNO-group and all nuclei obtained in frames of LoD, LoRD, NoD and NoRD models. Relativistic speed limit requirement taken into account in LoRD and NoRD leads to steepening of spectra as mentioned above. The energy of the `knee' $E_\mathrm{knee}$ obtained in NoRD is very close to provided in \cite{Blasi1} by exponential cut-off of injection spectrum, but we observe power law NoRD-spectrum above $E_\mathrm{knee}$ in contrast to exponential decay obtained in~\cite{Blasi1}. Sure, this requires greater maximum energies of particles accelerated in sources. The exponentially falling tails for all nuclei contradict to observed power law spectrum above $E_\mathrm{knee}$, if we consider it as produced by GCRs (without extragalactic component).
From this point of view NoRD-model seems to be preferable.

\begin{acknowledgments}
The reported study was supported by Russian Foundation for Basic Research (project~16-01-00556).
\end{acknowledgments}

\bigskip 

\end{document}